\documentclass[prx, amsmath, amssymb, twocolumn, nofootinbib]{revtex4-2}
\usepackage{dcolumn}
\usepackage{bm}
\usepackage{color}
\usepackage{graphicx}
\usepackage{algorithm}
\usepackage[noend]{algpseudocode} 
\usepackage{caption}

\usepackage[bitstream-charter]{mathdesign}

\definecolor{commentgray}{rgb}{0.5,0.5,0.5} 

\usepackage{subcaption}
\usepackage[justification=RaggedRight]{caption}

\algnewcommand{\LineComment}[1]{\State \(\triangleright\)\,\,\textit{#1}} 
\algnewcommand{\LineCommentOut}[2]{\\\hspace*{#2}\(\triangleright\)\,\,\textit{#1}}
\algnewcommand{\IIf}[1]{\algorithmicif\ #1\ \space}
\algnewcommand{\ElseIIf}{\unskip \algorithmicelse \space}
\algnewcommand{\EndIIf}{\unskip\ \algorithmicend\ \algorithmicif}

\usepackage[normalem]{ulem}

\def\pdisk{{\mathbb{D}_2}}

\makeatletter
\newcounter{phase}[algorithm]
\newlength{\phaserulewidth}
\newcommand{\setphaserulewidth}{\setlength{\phaserulewidth}}

\makeatother
\setphaserulewidth{.7pt}

\begin{document}

\preprint{AAPM/123-QED}

\title[Hyperbolic Tiling Neighborhoods in O(1) time]{\Large{\vspace*{3mm} Hyperbolic Tiling Neighborhoods in O(1) time}\\\vspace*{5mm}}

\author{Yanick Thurn}
\email{yanick.thurn@uni-wuerzburg.de}
\affiliation{ 
Julius-Maximilians-Universität W\"urzburg (JMU), \\Institute for Theoretical Physics and Astrophysics, W\"urzburg, Germany}

\author{Manuel Schrauth}
\email{manuel.schrauth@iis.fraunhofer.de}
\affiliation{Fraunhofer Institute for Integrated Circuits (IIS), Erlangen, Germany}
\affiliation{ 
	Julius-Maximilians-Universität W\"urzburg (JMU), \\Institute for Theoretical Physics and Astrophysics, W\"urzburg, Germany}
	
\author{Johanna Erdmenger}
\affiliation{ 
Julius-Maximilians-Universität W\"urzburg (JMU), \\Institute for Theoretical Physics and Astrophysics, W\"urzburg, Germany }

\date{\today}

\begin{abstract}
Tilings of the hyperbolic plane are of significant interest among many branches of mathematics, physics and computer science. Yet, their construction remains a non-trivial task. Current approaches primarily use tree-based recursive algorithms, which are fundamentally limited: they do not readily yield the neighborhood graph representing cell adjacencies, which is however required for many applications. We introduce a novel approach that allows to build hyperbolic tilings and their associated graph structure simultaneously, using only combinatoric rules without requiring an explicit coordinate representation. This allows to generate arbitrarily large, exact hyperbolic graphs, with an algorithmic complexity that does not depend on the lattice size. We provide an easy-to-use implementation which substantially outperforms existing methods, hence rendering ultra large-scale numerical simulations on these geometric structures accessible for the scientific community.
\end{abstract}

\maketitle
\section{Introduction}

In recent years, tilings of hyperbolic spaces emerged as a compelling geometric framework for exploring systems with inherent negative curvature. Applications range from critical phenomena~\cite{sakaniwa2009,iharagi2010,gendiar2014,baek2007,baek2009b,baek2009c,baek2009,baek2010,lopez2017} to condensed matter physics \cite{Bienias:2021lem}, topolectric circuits \cite{PhysRevLett.125.053901,Lenggenhager2021,Chen:2022}, circuit quantum electrodynamics \cite{Kollar2019,Kollar,Boettcher:2019xwl}, quantum field theory \cite{brower2021,Brower:2022atv},  and AdS/CFT correspondence \cite{Boyle2020,Boyle2025,basteiro2022,basteiro2023,Erdmenger2025}, to name only a few. It is clear that the steadily growing significance of hyperbolic geometry~\cite{coxeter1989} requires robust algorithms for generating these structures, ideally accompanied by modern, high-performane implementations. This has led to the development of library projects such as \emph{hypertiling}~\cite{hypertiling,schrauth2023}.

A common way of constructing tessellations that exhibit a certain degree of regularity is to start with a fundamental (seed) cell which is either reflected across its edges or rotated around its vertices in order to generate further offsprings~\cite{humphreys1990}. This kaleidoscopic method is quite general and can be used for various cell shapes~\cite{Kaplan2000}, from polygons in 2D space to general polytopes in higher dimensions. In order to create a tiling / tessellation that populates the underlying manifold without voids or overlaps, certain geometrical restrictions must be met~\cite{coxeter_1997}, as there is generally no concept of similarity in curved manifolds. A very fundamental and frequently used seed shape is the Schwarz triangle, defined by a triple of integers $(p,q,r)$ which indicate the interior vertex angles $\pi/p$, $\pi/q$ and $\pi/r$ of the triangle and hence the number of adjacent cells sharing a vertex, namely $2p$, $2q$ or $2r$, respectively. If $1/p+1/q+1/r<1$, the tessellation is genuinely hyperbolic~\cite{poincare1882}, with a sum of inner angles of less than $\pi$. By concatenating adjacent triangles appropriately, e.g.~the $2r$ triangles around an $r$-vertex, regular \textit{polygonal} tilings can be constructed. Those are formally labeled by the Schläfli symbol $(p,q)$, where $p$ denotes the number of edges of the $p$-gon and $q$ the number of adjacent polygons at its vertices.

\begin{figure*}[t]
	\begin{subfigure}{0.33\linewidth}
		\includegraphics[width=.9\linewidth]{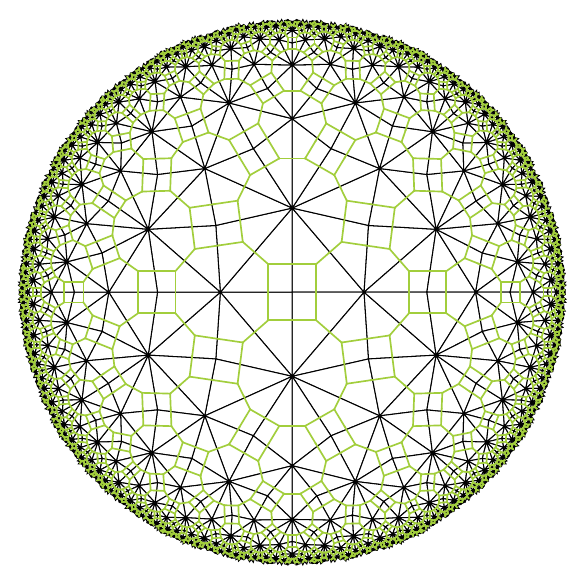}
		\caption{(5, 4, 2)}
	\end{subfigure}%
	\begin{subfigure}{.33\linewidth}
		\includegraphics[width=.9\linewidth]{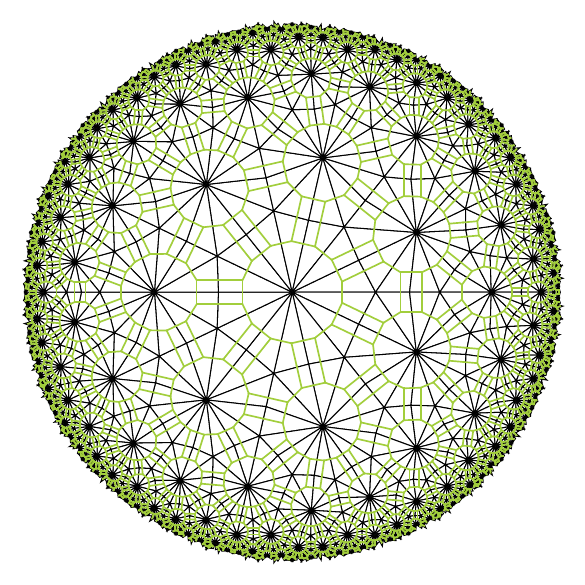}
		\caption{(2, 3, 7)}
	\end{subfigure}
	\begin{subfigure}{.33\linewidth}
		\includegraphics[width=.9\linewidth]{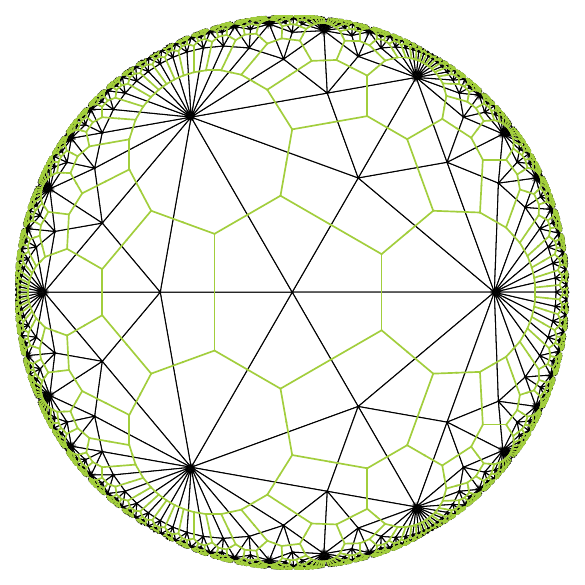}
		\caption{(3, 9, 3)}
	\end{subfigure}
	\caption{Three hyperbolic triangle tilings (black) and their neighbor relations (green), embedded in the Poincaré disk model.}
	\label{fig:fancy_lattices}
\end{figure*}

A particular challenge lies in the representation of coordinates. Several established models exist, among which the Poincaré disk model is particularly popular as it provides conformal mappings, compactness, and a high degree of symmetry. However, it is within the nature of the kaleidoscopic method that an accumulation of numerical errors through repeated transformations can not be avoided -- regardless of the choice of coordinates~\cite{celinska2024}. 

Formalizing the construction of a tiling in a purely combinatorial manner avoids the requirement for coordinates entirely. These approaches have a long history in mathematical literature~\cite{coxeter1954} and make particular use of the theory of triangle reflection groups $\Delta(a,b,c) \subset \mathrm{SL}_2(\mathbb{R})$, which feature three generators $a$, $b$, $c$, associated to triangle edge reflections. They serve the conditions $ a^2+b^2+c^2 = 1 $ and  $ (ab)^p = (bc)^q = (ca)^r = 1 $, where integers $p$, $q$, $r$ are defined as above, yielding a formal definition of a triangular tiling. Generalizing this concept gives rise to the theory of Coxeter groups~\cite{coxeter1989}, which extend beyond triangular configurations to encompass polygonal and, in higher dimensions, polyhedral shapes~\cite{humphreys1990}. A geometric method which makes use of the symmetries of Coxeter groups to construct tessellations from regular polytopes is the so-called Wythoff construction. The tessellation is fully determined by the seed shape and the selected symmetry group. Specifically, one encounters what is called a word problem, which involves deciding whether two words represent the same element in a group. For Coxeter groups, given their definition through chains of generators, the word problem becomes a matter of string manipulation according to specific rules. This directly ties into the concepts of formal languages, where words and manipulation rules (grammars) are fundamental and it features significant overlap with the theory of automata. In particular, finite state automata can be used to recognize reducible sequences of reflections and implement them numerically~\cite{casselman1995,casselman2002,casselman2008}. There exists a number of algorithmic implementations for hyperbolic tilings, which are inspired by the Wythoff principle, tracing back to the pioneering work by Dunham~\cite{dunham1981,dunham1986,dunham2007} in the early 1980s and including more recent approaches such as \cite{kopczynska2021,kopczynska2022b}, as well as game-based procedural construction engines, such as beautifully demonstrated in Reference~\cite{kopczy2024}. 

The approaches mentioned above construct tilings as hierarchical tree structures. Even though conceptually very elegant, this depth-first approach is not readily capable of obtaining adjacency/neighborhood relations among cells or compute shortest paths between two vertices, as nearby cells can end up in entirely different branches of the tree. 
Hence, even though Wythoff-inspired combinatorial algorithms are very successful in constructing the tiling (i.e.~the set of cells), the associated graph structure is generally not discussed in most of the corresponding literature. A notable exception is the work by M.~Margenstern~\cite{margenstern2000, margenstern2006, margenstern2010}, who, in a series of papers, showed how particular polygonal tilings (such as (5,4) and (7,3)) can be constructed by a so-called splitting method. This process allows to encode positions of vertices in the tree uniquely as maximal Fibonacci sequences. By traversing this tree in a separate processing step, this unique coordinate-like representation allows to locate tiles, calculate distances and hence reconstruct the cell neighborhood.

In this manuscript we introduce a novel algorithm for the construction of tilings in the hyperbolic plane, featuring significant advancements over existing methods. We achieve a modular design, decoupling cell propagation, coordinate management, and construction of the graph structure, while retaining the kaleidoscopic principle. This is accomplished through a loop-based approach and a formalization of cell types based on local vertex configurations, hence fully eliminating the need for complex branching or tree structures. As a result, our method features an exact, coordinate-free construction while delivering substantial new advantages.

First and most importantly, adjacency relations, specifically the so-called von Neumann neighborhood of cells is inherently generated during construction of the tiling with no additional dependence on the lattice size or layer depth, i.e.~in $\mathcal{O}(1)$ time. To the best of our knowledge, this has not been achieved before. 
Tree-based algorithms typically require a separate processing step to reconstruct the full neighborhood of a vertex, as the challenge is to properly interconnect individual branches. We consider the splitting method~\cite{margenstern2000} as state-of-the-art in terms of computational complexity, where, for a number of particular polygonal tilings it was shown that a time complexity of $\mathcal{O}(n)$ can be achieved, where $n$ is the number of generated layers (i.e. the tree depth), or, equivalently, in $\mathcal{O}(\log(N))$ time, where $N$ is the number of cells in the tiling.

Second, compared to other approaches~\cite{dunham2007, margenstern2006}, we consistently support all combinations of regular polygonal and Schwarz triangle cells, with the modular design allowing for optional coordinate embedding using any suitable coordinate system.

Third, our approach is particularly lightweight and computationally efficient, as, for instance, recursion and exponential functions can be fully avoided -- in favor of integer arithmetics and if-statements. The algorithm operates locally, using only vertex states and polygon types, and therefore does not require to build and maintain sequences of transformations. A performance analysis demonstrates that our method consistently outperforms existing methods, likely making it the fastest for tilings of the hyperbolic plane. In particular, we provide a Python implementation along with a user-friendly wrapper for the \emph{hypertiling} library\footnote{compare \emph{www.hypertiling.de} or the package \emph{hypertiling} on the PyPI package index}, optionally accelerated with numba JIT compilation.

The remainder of this article is structured as follows: In Section~\ref{sec:GeneralConcept} we lay out the general tiling construction mechanics and discuss how neighbor relations are deduced. Section~\ref{sec:Performance} provides benchmarks in terms of computational efficiency, before we conclude in Section~\ref{sec:Conclusion}. A detailed, albeit more technical description of our approach is available in the Appendix.
\section{General Concept}
\label{sec:GeneralConcept}

Hyperbolic tessellations are usually constructed by repeated transformations starting from a single seed region with suitable geometric shape, marking the first layer. We employ reflections across open (i.e.~vacant) edges of existing cells, in contrast to, e.g., rotations around vacant vertices~\cite{dunham1981,dunham1986,dunham2007}. Open edges are those which do not have a partner cell on the other side yet. A new layer is obtained by applying reflections on all currently open edges of the previous layer, as described in the code fragment in Fig.~\ref{alg:main}. When using coordinates, such as for example the hyperbolic Poincaré disk, reflections are implemented as a series of Moebius transformations in the complex plane, in particular a combination of rotation, translation, conjugation and inverse rotation~\cite{hypertiling}. 

\begin{figure}[t!]
	\hrule height \phaserulewidth \vspace{3mm}
\begin{algorithmic}
	\State \textcolor{commentgray}{\LineComment{Initialize the first layer manually}}
	\State \emph{current\_layer} = \textsc{construct\_seed\_region()}{}
	\State \textcolor{commentgray}{\LineComment{Construct tiling and graph}}
	\For{i = 1 \textbf{to} \emph{number\_of\_layers-1}} \textcolor{commentgray}{\Comment{repeat until desired size}}
	
	\State \emph{next\_layer} $\leftarrow$ [] \textcolor{commentgray}{\Comment{prepare container}}
	
	\For{\emph{cell} \textbf{in} \emph{current\_layer}}
	\For{\emph{vertex} \textbf{in} \emph{cell}}
	\State \emph{child} $\leftarrow$ \textsc{propagate}{(\emph{cell}, \emph{vertex})} \textcolor{commentgray}{\Comment{core mechanism}}
	\State \textsc{graph}{(\emph{child}, \emph{cell})} \textcolor{commentgray}{\Comment{establish graph}}
	\State \textsc{coordinates}{(\emph{child})} \textcolor{commentgray}{\Comment{migrate coordinates}}\\
	
	\State \emph{next\_layer}.\Call{add}{\emph{child}}
	\EndFor
	\EndFor
	\State \emph{current\_layer} $\leftarrow$ \emph{next\_layer}
	\EndFor
\end{algorithmic}
	\vspace{3mm}\hrule height \phaserulewidth \vspace{2mm}
	\caption{Pseudo code of main method highlighting the layer-wise construction scheme where propagation, graph structure and coordinates are decoupled. A detailed discussion of \textsc{propagate} and \textsc{graph} functions can be found in Appendix~\ref{sec:AlgorithmicDetails}. }
	\label{alg:main}
\end{figure}

It is useful to employ the terms \emph{parent} and \emph{child} to denote the hierarchical relationship in offspring-based cell generation. Additionally, we define \emph{co-parent} an any additional cell in the previous layer other than the actual parent that shares an edge with the child.

In this article we focus on tilings by Schwarz triangles, denoted by a triplet $(p,q,r)$, as this is the more general case compared to regular polygons. As detailed earlier, regular $(p,q)$ tilings can be constructed by gluing together any $2r$ cells sharing an $r$ vertex. Additionally, in Appendix~\ref{sec:RegularPolygons}, we adjust our approach to obtain regular tilings directly, with an accompanying implementation also available.

\begin{figure}[t]
	\centering
	\includegraphics[width=0.85\linewidth]{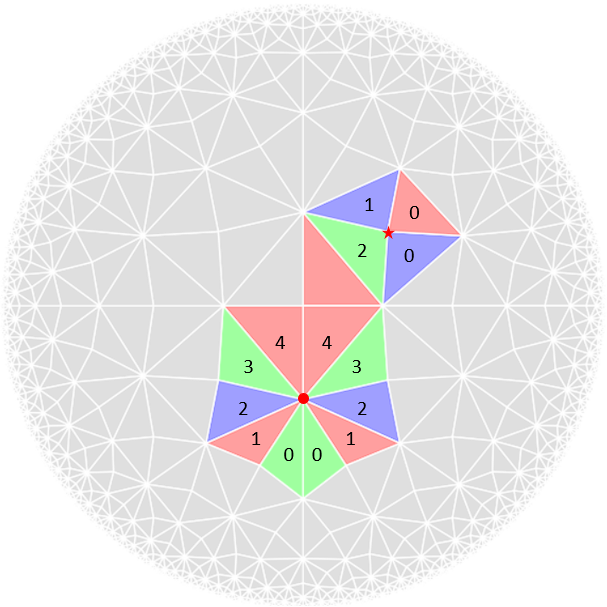}
	\caption{Illustration of how vertices are closed by asymmetric (star vertex) and symmetric (red dot vertex) filler triangles in a (5,4,2) tiling. Labels correspond to the respective valence counter, i.e~the number of additional triangles required to close the vertex.}
	\label{fig:filler_types}
\end{figure}

Fundamentally, for every triangle in the tiling a \textit{valence counter} is assigned to each vertex, which is decremented as new triangles are generated around that vertex. Once a counter reaches zero, the vertex is considered \emph{closed} and the generation of further offsprings is terminated. The specific type of vertex ($p$, $q$, or $r$) directly sets the initial value of its associated counter. As an example, in Figure~\ref{fig:filler_types}, we consider the vertex marked by a red dot, which is introduced by one of the two central red polygons with label 4. The number of further offspring cells required to close this vertex is decremented as new triangles are generated within subsequent layers.

When performing an edge reflection, the newly generated cell naturally shares two vertices with its parent. Regarding the third vertex, we distinguish two cases:
\begin{itemize}
	\item The vertex is newly introduced. Then, the counter is initialized as $(x-1)$, with  $x\in{p,q,r}$.
	\item The vertex already exists in the tiling. In this case the initial value of the counter is modified according to the current situation at this vertex.
\end{itemize}
As an example, the red-star vertex in Figure~\ref{fig:filler_types} is opened by two green triangle with label 2 and closed by the red triangle with label 0.

Even though creating a tiling by successive reflections may appear straightforward, two important points need to be considered. When implemented in a naive fashion, successive reflections will in general generate duplicates~\cite{hypertiling}. Moreover, adjacency relations between different propagation branches can not be deduced straightforwardly. We resolve this by establishing a notion of \textit{filler} cells:
\begin{itemize}
    \item \emph{Asymmetric filler triangles} have one parent and one co-parent. The remaining edge is connected to a child. The name stems from the fact, that these type of triangles close a vertex asymmetrically. They do not open a new vertex. An example is given by the red triangle with label $0$ in Figure~\ref{fig:filler_types}). 
    
    \item \emph{Symmetric filler triangles} have one parent, one child and one symmetrical partner, located in the same layer. The pair of two triangles closes the vertex symmetrically. Symmetric fillers, together with their partner, always open a new vertex, which is closed symmetrically. Examples are the two green triangles with label $0$ in Figure~\ref{fig:filler_types}). 
    
    \item \emph{Regular triangles} have one parent, blocking one edge. The two remaining edges connect to two children. Upon construction, triangles of this type do not close any vertex.
\end{itemize}
Note that even though uniform tilings are vertex-transitive, this classification depends on the seed position and the order of their creation. From the algorithmic perspective, correctly identifying the type of a newly generated triangle and initializing its vertex valence counters appropriately is the central challenge of our approach and is discussed in the following.

Children are created by reflection on up to two edges open by the parent. Each edge connects two vertices. 
An edge is blocked if one of the corresponding counters is zero. As long as the values of both counters are greater than zero, more offsprings are created, decrementing the number of cells to close the vertex by one each.

In case the parent is a regular triangle, the non-participating vertex opens a new vertex, which will eventually be closed by an asymmetric filler type. This can be understood by the fact that each vertex requires either $2p$, $2q$ or $2r$ triangles to be closed -- hence if the vertex is opened by a single triangle this leads to an asymmetric closing. As a consequence, in order to prevent over-counting, the counter clockwise path is (by choice) decremented twice, initially. This way, the vertex is closed by an asymmetric filler cell originating from the opposite (clockwise) direction.

\begin{figure}[t]
	\begin{subfigure}{0.33\linewidth}
		\includegraphics[width=\linewidth]{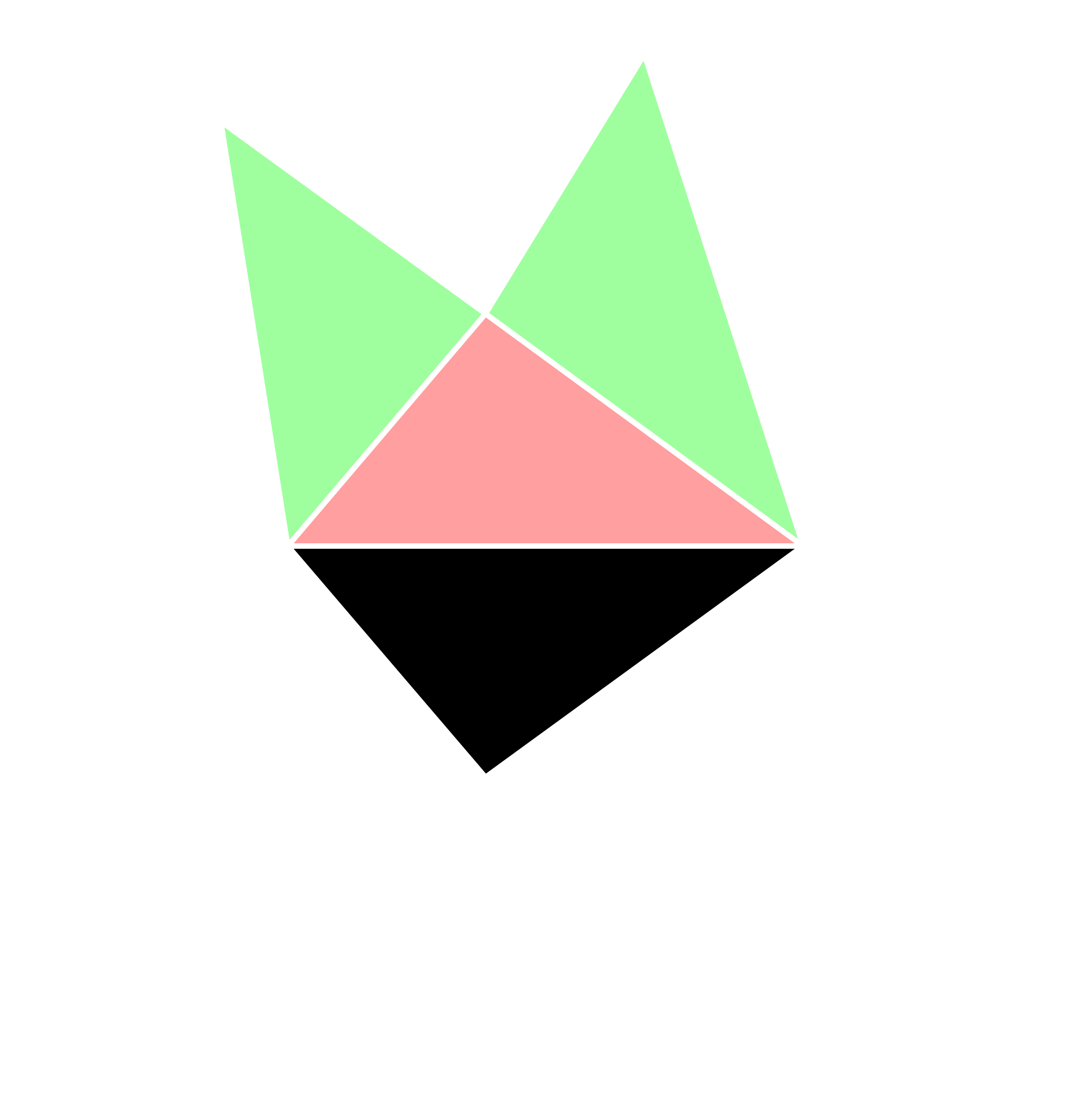}
		\caption{Regular triangle}
		\label{fig:sfig1}
	\end{subfigure}%
	\begin{subfigure}{.33\linewidth}
		\includegraphics[width=\linewidth]{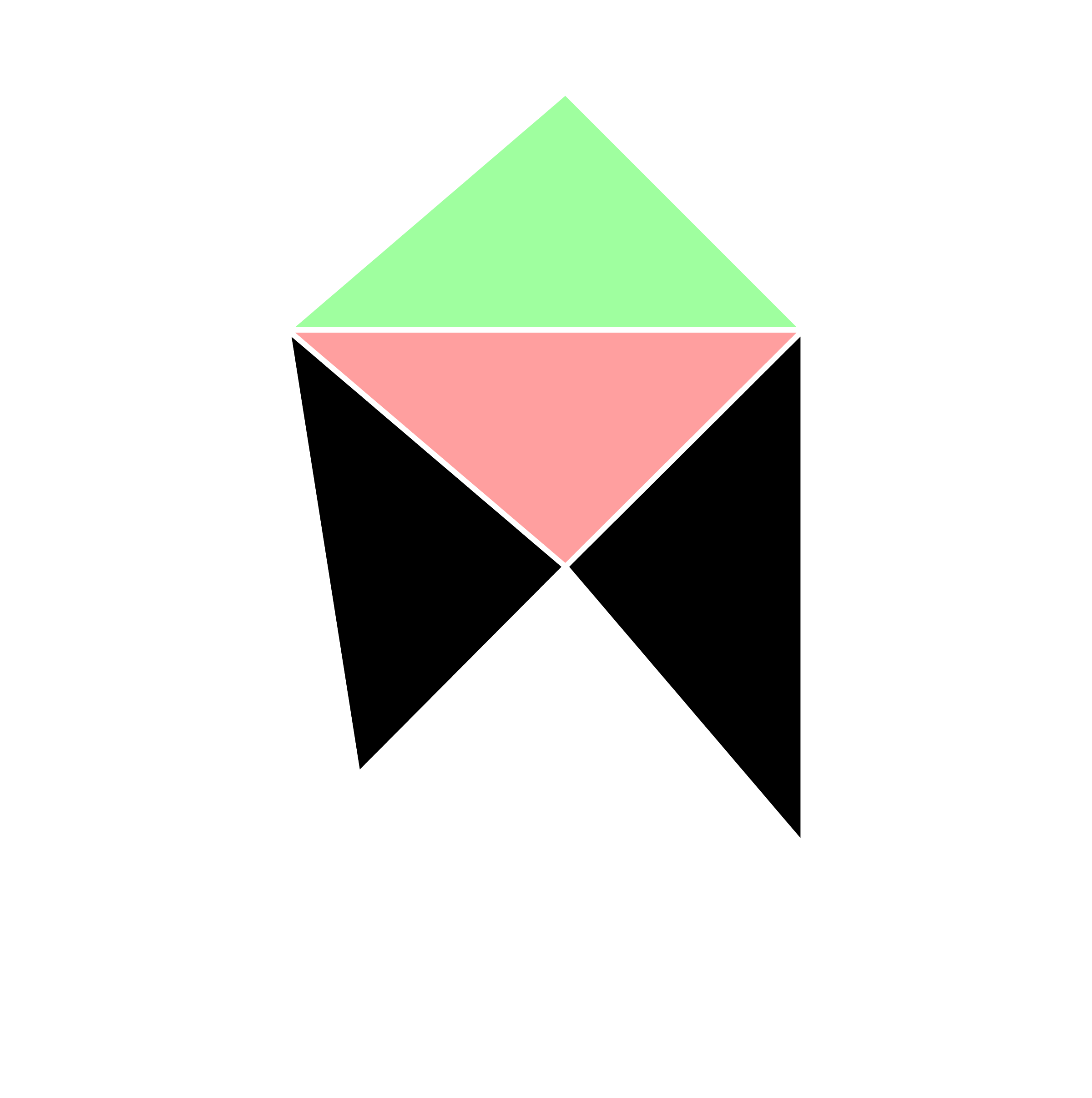}
		\caption{Asymmetric filler}
		\label{fig:sfig2}
	\end{subfigure}
	\begin{subfigure}{.33\linewidth}
		\includegraphics[width=\linewidth]{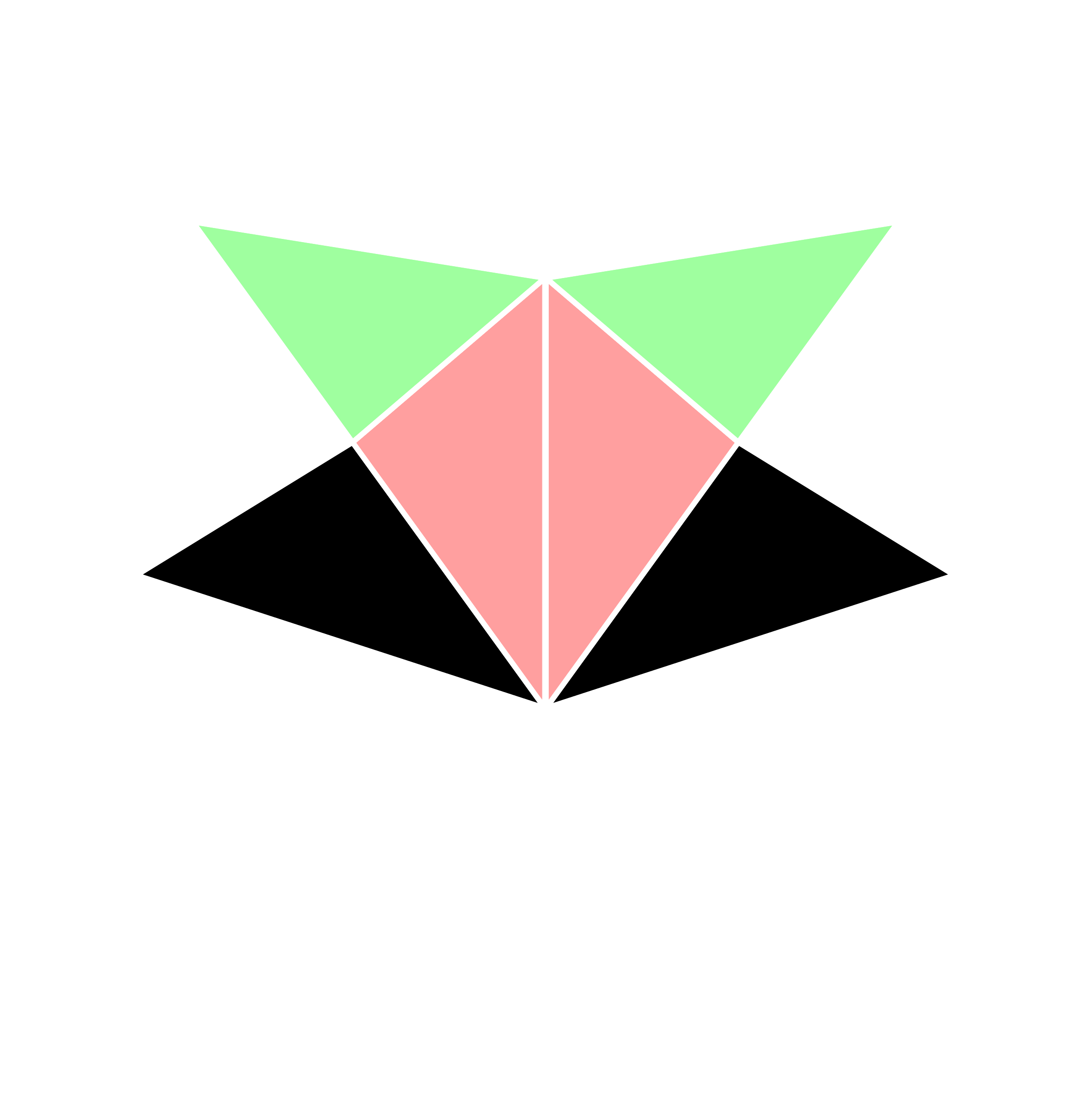}
		\caption{Symmetric filler }
		\label{fig:sfig3}
	\end{subfigure}
	\caption{Illustration of triangle types, where black triangles denote the parent layer and green ones denote the children. Only for regular triangles (a), two edges connect to children.	}
	\label{fig:triangle_types}
\end{figure}

Neighbor relationships among cells define the edges of the associated hyperbolic \emph{graph} (green lines in Figure~\ref{fig:fancy_lattices}). Assigning coordinates to the vertices of this graph, e.g.~located at the centers of the corresponding tiles, yields what is known as the dual tiling. 
As already hinted in Figure~\ref{alg:main}, the corresponding method executes after the cell propagation, yielding the graph in only $\mathcal{O}(1)$ additional time, therefore independent of the current tiling size. Detailed pseudo code for both methods (\textsc{propagate} and  \textsc{graph}) can be found in Appendix~\ref{sec:AlgorithmicDetails}.

Regular triangles have no additional neighbors beyond their parent and children. Symmetric filler triangles always appear in adjacent pairs, as detailed above. Consequently, the triangles in such a pair are mutual neighbors. Asymmetric filler triangles can be interpreted as originating from \emph{two} triangles in the previous generation. While only one is the actual parent, the other contributes as an additional neighbor (compare $\star$ vertex in Figure~\ref{fig:filler_types}). This represents the only instance where a neighbor (other than the parent) resides in a previous layer.  
As an example, in Figure~\ref{fig:filler_types}, the red triangle with counter $0$ is generated from the blue triangle with counter $1$, its parent. However, the blue triangle with counter $0$ is also a neighbor of the red triangle, despite not being its parent and belonging to the previous generation. 
\section{Performance}
\label{sec:Performance}

We provide an open source Python implementation, with optional Numba just-in-time (JIT) compilation, accessible through the \emph{hypertiling} package. Specifically, two new tiling kernels are available: Generative Reflection Combinatoric (GRC) for polygonal tilings and GRC-Triangle (GRC-T) for general Schwarzian triangle tilings. Moreover, a C++ implementation of GRC, which exhibits comparable execution speeds to the Numba-optimized Python version, while further improving memory efficiency through the utilization of 8-bit integer data types for relevant internal representations, will be available shortly.

In this section we the average asymptotic cell construction time results for both kernels. The memory footprint is briefly discussed in Appendix~\ref{sec:AppendixMemory}. All measurements were performed on an Intel Xeon w3-2435 operating at 3.1 GHz, with 64 GB of RAM, under a Debian Linux environment and we use Poincaré disk coordinates.

\begin{figure}[t!]
	\includegraphics[width=\linewidth]{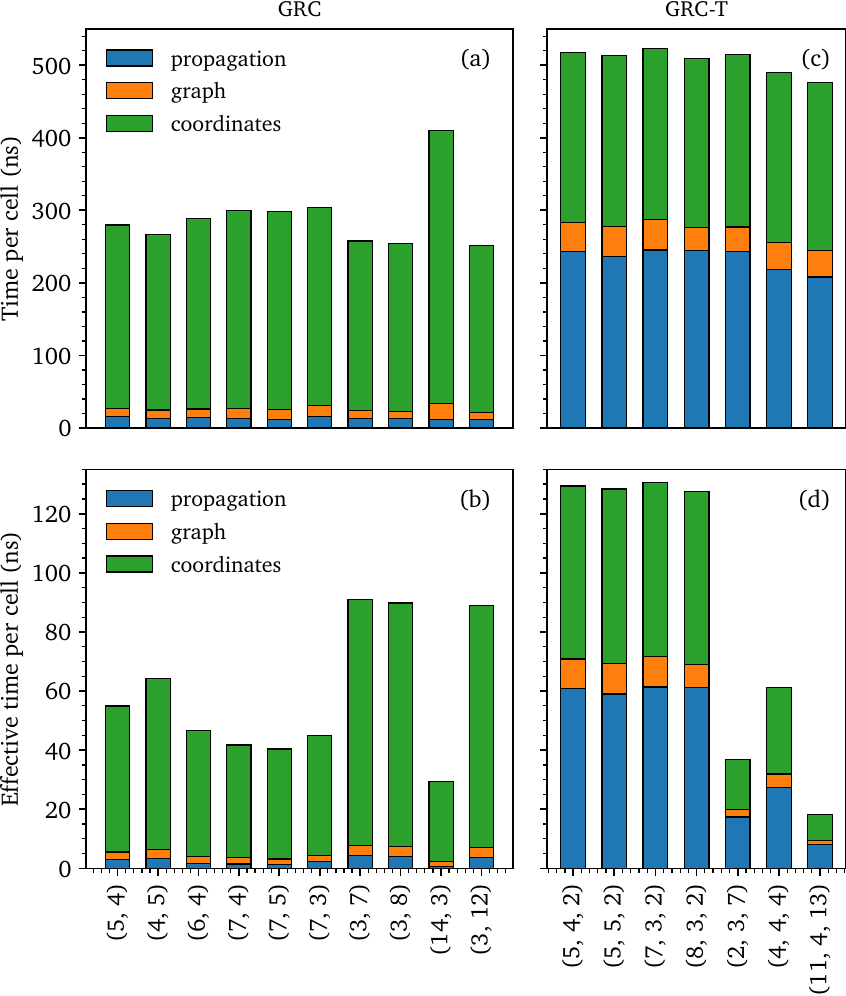}
	\caption{Upper row: average construction time per cell for polygonal tilings (GRC) and triangle tilings (GRC-T). Lower row: corresponding effective time per cell when only the fundamental symmetry sector is explicitly constructed, with the remaining cells generated on-demand via symmetry operations. Colors relate to processing stages.}
	\label{fig:bench01}
\end{figure}

\paragraph{Polygonal tilings}

Figure~\ref{fig:bench01} (left column) presents the GRC kernel results, with corresponding figures provides in Table~\ref{tab:benchgrc}. The upper panel (a) illustrates the \emph{true} time per cell, whereas in the lower panel, we leverage the discrete $p$-symmetry and construct only a fundamental sector, therefore reducing the \emph{effective} construction time. Through the generator functionality (compare~\cite{schrauth2023}), this can be implemented efficiently with cells outside the sector being generated on-demand. As a result, the performance is primarily influenced by the $p$ symmetry, hence the effective time required per polygon reduces as the relative size of the fundamental sector decreases.

From the figure, it becomes clear that, overall, the computational cost is dominated by (optional) coordinate calculations. This can be attributed to two main factors. Firstly, the reflection operation is applied to each vertex, resulting in $p$ operations per polygon. 
Secondly, explicit coordinate transformations (in this case Moebius functions) typically involve sequences of complex multiplications, divisions, and trigonometric function evaluations. In contrast, the cell propagation and graph determination processes (blue and orange bars, respectively) primarily involve basic integer arithmetics.

\begin{table*}
	\renewcommand{\arraystretch}{1.1} 
	\begin{tabular}{c|cccc|c}
						Kernel &GRC &&&& DUNX \\
		\hline\hline
		$(p,q)$        & \phantom{m}\textsc{Propagate} (\textsc{P})\phantom{m} &  \phantom{mm}\textsc{Graph}  (\textsc{G}) \phantom{mm} & \textsc{Coordinates} (\textsc{C}) &\phantom{mmm} \textsc{P+C}\phantom{mmm}      & \phantom{mmm}\textsc{P+C}\phantom{mmm}  \\ \hline
		\hline
		\textbf{5, 4}  & $15.91 \pm 0.74$                & $11.06 \pm 0.90$                  & $252.8 \pm 4.1$                   & $268.7 \pm 4.2$   & $684.7 \pm 3.7$ \\
		\textbf{4, 5}  & $13.38 \pm 0.27$                & $11.42 \pm 0.93$                  & $241.9 \pm 4.1$                   & $255.3 \pm 4.1$   & $662.6 \pm 4.9$ \\
		\textbf{6, 4}  & $13.92 \pm 0.56$                & $12.09 \pm 0.62$                  & $263.1 \pm 6.2$                   & $277.0 \pm 6.2$   & $677.0 \pm 5.5$ \\
		\textbf{7, 4}  & $13.23 \pm 0.65$                & $13.7 \pm 1.4$                    & $272.8 \pm 4.5$                   & $286.1 \pm 4.6$   & $667.6 \pm 6.0$\\
		\textbf{7, 5}  & $11.68 \pm 0.31$                & $14.2 \pm 1.0$                    & $272.1 \pm 5.7$                   & $283.8 \pm 5.7$   & $619.3 \pm 3.5$\\
		\textbf{7, 3}  & $16.18 \pm 0.65$                & $14.71 \pm 0.83$                  & $273.3 \pm 4.8$                   & $289.5 \pm 4.8$   & n.a.\\
		\textbf{3, 7}  & $13.45 \pm 0.39$                & $10.83 \pm 0.99$                  & $233.3 \pm 3.8$                   & $246.8 \pm 3.9$   & n.a.\\
		\textbf{3, 8}  & $12.84 \pm 0.32$                & $9.60 \pm 0.82$                   & $231.4 \pm 3.7$                   & $244.2 \pm 3.7$   & n.a.\\
		\textbf{14, 3} & $12.10 \pm 0.36$                & $21.9 \pm 1.4$                    & $375.7 \pm 5.5$                   & $387.8 \pm 5.5$   & n.a.\\
		\textbf{3, 12} & $11.64 \pm 0.44$                & $9.86 \pm 0.96$                   & $230.5 \pm 3.8$                   & $242.1 \pm 3.8$   & n.a.\\
	\end{tabular}
	\caption{Comparison of timing results per cell, in nanoseconds, for the GRC construction kernel (developed in this manuscript) and the established method by Dunham (DUNX), both implemented in the \emph{hypertiling} package. Generally, for GRC, the construction of only one symmetry sector is sufficient, hence the speed advantage compared to DUNX is more than one order of magnitude for actual applications. The graph structure is not available for DUNX.}
	\label{tab:benchgrc}
\end{table*}

Timing results for the kaleidoscopic propagation (blue bars) and the adjacency reconstruction (orange bars) are of the same order of magnitude, with the latter generally exhibiting lower values for smaller $q$. This is attributed to the fact that the flag attributes of child polygons, as determined by the \textsc{propagate} method, inherently encode neighboring relationships. Consequently, explicit derivation of adjacency information primarily involves comparisons to ascertain the polygon type and structuring the neighboring indices into an appropriate data container. For larger values of $p$, the time required for neighbor determination can exceed that of the propagation process, as observed for the $(14,3)$ tiling. This is due to the greater number of neighbors, necessitating more computational operations, whereas in the propagation only two vertices need to manipulated, independent of $q$ (compare Appendix~\ref{sec:RegularPolygons}).

For the propagation function (compare Table~\ref{tab:benchgrc}), smaller values of $p$ correlate with increased processing time, as evidenced by examining sets of tilings with fixed $q$, such as $(5,4)$, $(6,4)$ and $(7,4)$. This can be explained by the more complex algorithmic handling required for the first and last child polygons compared to the middle ones (compare Appendix~\ref{sec:RegularPolygons}). 
In the case of $p = 3$, both generated child polygons are either the first or the last. For asymmetric filler polygons, a single child polygon serves as both the first and the last. Secondly, the propagation time demonstrates a slight dependence on $q$, which is apparent when comparing $(7,3)$, $(7,4)$ and $(7,5)$. This dependence arises from the fact that a larger value of $q$ reduces the relative amount of filler polygons, which are generally more complex to process than regular polygons.

It is important to note that the values presented in both upper panels are for comparative reference only, since our implementation, as previously stated, only necessitates the construction of a single symmetry sector. The corresponding timing results per polygon with respect to the full lattice are significantly lower, as shown in the lower panels, approximately scaled down by a factor of $p$. Consequently, while $(14,3)$ exhibits the highest computational cost when considering the full lattice construction, it becomes the least expensive among the presented tilings in the left part of the figure when only the fundamental sector is constructed.

\paragraph{General Triangle Tilings}

For the GRC-T implementation, results are presented in the right column of Figure~\ref{fig:bench01}, with detailed numbers provided in Table~\ref{tab:times_grct}. The overall performance when constructing only the fundamental sector (panel d) is primarily governed by the $2r$ sector symmetry. 
A particularly illustrative comparison involves the $(7, 3, 2)$ and $(2, 3, 7)$ tilings: though topologically equivalent, their fundamental symmetry sectors differ significantly in size, constituting $1/4$ and $1/14$ of the full tiling, respectively. 

In contrast to the GRC kernel, the propagation process (blue bars) in GRC-T is notably slower. This arises from the necessity to store and process all three vertices of each triangle as opposed to only the leftmost and rightmost vertices in GRC. Additionally, higher-order connections, specifically those involving the co-parent, need to be updated appropriately which does not occur in regular tilings due to their higher degree of symmetry. Generally, similar to GRC, larger values for any element of the triplet $(p, q, r)$ tend to improve the speed of triangle creation, as the relative number of filler triangles decreases.

Finally, the graph function (orange bars) remains relatively fast, as, similar to GRC, pre-determined vertex attributes encode the neighboring relationships, and the function, from a complexity standpoint, primarily performs the insertion of the correct indices into the corresponding adjacency data container.

\begin{table}[b]
\renewcommand{\arraystretch}{1.1} 
\begin{tabular}{c|ccc}
	$(p,q,r)$                      & \textsc{Propagate} (\textsc{P}) & \textsc{Graph}  (\textsc{G}) & \textsc{Coordinates} (\textsc{C}) \\ \hline\hline
	(5, 4, 2)                      & $243.7 \pm 5.8$                 & $39.9 \pm 8.5$                    & $234 \pm 11$                    \\
	(5, 5, 2)                      & $236.1 \pm 4.7$                 & $41.6 \pm 8.6$                    & $235.8 \pm 9.4$                 \\
	(7, 3, 2)                      & $245.5 \pm 5.6$                 & $41.8 \pm 9.0$                    & $235 \pm 12$                    \\
	(8, 3, 2)                      & $244.5 \pm 4.8$                 & $31.8 \pm 5.1$                    & $233.3 \pm 9.8$                 \\
	(2, 3, 7)                      & $244 \pm 4.9$                   & $33.6 \pm 7.9$                    & $237.3 \pm 9.9$                 \\
	(4, 4, 4)                      & $218.4 \pm 4.9$                 & $37.0 \pm 7.3$                    & $234 \pm 10$                    \\
	(11, 4, 13)                    & $207.9 \pm 4.0$                 & $36.5 \pm 7.7$                    & $231.5 \pm 8.6$                 \\ 
\end{tabular}
\caption{Timing results for the GRC-T construction kernel, per triangle, in nanoseconds. Generally, building only one symmetry sector is sufficient, lowering the time required  effectively by a factor of $2r$ with respect to the total number of cells.}
\label{tab:times_grct}
\end{table}

\section{Conclusion}
\label{sec:Conclusion}

This work presents a novel algorithm for generating arbitrary triangular and polygonal hyperbolic tilings and their associated graphs. Due to its purely combinatorial nature, an explicit coordinate representation is not required. 
This is achieved through a non-recursive loop-based design where the construction is performed iteratively layer by layer, embedding the necessary propagation information locally within each cell, thus avoiding hierarchical branching. A significant advantage of this approach lies in its ability to determine cell adjacency relations concurrently with the tiling generation, with no further effort in terms of computational complexity. The resulting hyperbolic graph is commonly required in scientific and technical applications which go beyond simple visualization or artistic purposes~\cite{ouyang2014,ouyang2019}. The modular design of our algorithm cleanly separates the core cell propagation logic from neighbor identification and optional coordinate embeddings. In particular, this allows for flexible integration of any suitable coordinate system, and furthermore, can be adapted for flat and spherical tilings. 

We provide highly optimized implementations as part of the open-source \emph{hypertiling} library as new kernels GRC (for polygonal tilings) and GRC-T (for general triangles). Both kernel feature high memory efficiency by exploiting the fundamental discrete symmetry of the tiling and constructing cells on demand via iterator functionality. To the best of our knowledge, this strategy is not available in other implementations of hyperbolic tiling algorithms. In particular, benchmarks against a technically equally optimized implementation of the well-established algorithm by Dunham~\cite{dunham2007} reveal a performance improvement of at least a factor of two (compare Table~\ref{tab:benchgrc}), even when the sector-symmetry is \emph{not} used, while natively covering more $(p,q)$ combinations and, as discussed, simultaneously providing adjacency information, which is not available for Dunham's hierarchical approach. Leveraging the sector-based construction further amplifies the speed advantage, exceeding one order of magnitude. Achieving a construction time of mere nanoseconds per cell (Table~\ref{tab:benchgrc}), our implementation approaches the practical limits of sequential optimization. Future work will include parallelization strategies to further enhance performance. Currently, we are also working on extending our approach to other hyperbolic structures, such as wormhole tilings and compactified tessellations, which are particularly relevant in solid state physics very recently~\cite{PhysRevLett.133.061603}.

\appendix

\section{Algorithmic Details}
\label{sec:AlgorithmicDetails}
In Figure~\ref{alg:propagate} we provide a detailed pseudo code for the function \textsc{propagate} which is the main driver of triangle replication. During the construction process, this function is invoked for every vertex of every triangle in the currently outmost layer. As the function contains no loops or recursive calls, the time complexity is trivially $\mathcal{O}(1)$, i.e.~new cells are constructed in a fixed amount of time, independent of the current tiling size. For every triangle, the following variables are stored
\begin{itemize}
	\item the final number of triangles per vertex as array \textit{edges} with entries that are permutations of $(p,q,r)$,
	\item three \textit{valence counters}, initialized as $p$, $q$ and $r$,
	\item a global \textit{id} $\in \mathbb{N}$,
	\item a set of flags  $ \in\{\mathtt{f}, \mathtt{F}, \mathrm{None}\}$, allowing to distinguish between the counter clockwise (i.e.~\emph{short}) branch around a vertex closed by an asymmetric filler cell ($\mathtt{f}$) and the clockwise branch  ($\mathtt{F}$). If the flag is $\mathrm{None}$, the vertex will be closed by a pair of symmetric filler cells instead,
	\item optionally, a set of coordinates, provided in suitable representation, e.g. $c_i \in \pdisk = \{z\in \mathbb{C}, \,|z|<1\}$
\end{itemize}

The construction process begins by manually placing the innermost layer, consisting of $2r$ triangles. Specifically, one fundamental triangle with an $r$-vertex at the intended center of the tiling needs to be generated. Subsequent reflections over the 'inner' edges 
close the central vertex by generating the remaining $2r-1$ triangles. 
As the $2r$ innermost triangles are all part of the same layer, no particular creation order and hence no classification as regular or filler triangles can be applied. As a result, for the second layer, the propagation function is used with a slight modification. Specifically, for each triangle, only a child from the edge connecting the $p$ and $q$ vertices, is created. Each subsequent layer is generated by iterating through triangles in the parent layer, and for each of their vertices invoking the propagation function, compare Figure~\ref{alg:propagate}. In the following we discuss this method in detail. 

\begin{figure}[t!]
	\hrule height \phaserulewidth \vspace{3mm}
	\begin{algorithmic}[5]
		\Function{PROPAGATE}{\emph{parent: Triangle}, \emph{$v_e$: int}}
		
		\LineComment{$v_e$, $v_a$ , $v_n$  = (edge, adjacent, non) participating vertex}
		\State \emph{$v_a$} $\leftarrow$ (\emph{$v_e$} + 2) mod 3
		\State \emph{$v_n$} $\leftarrow$ (\emph{$v_e$} + 1) mod 3
		\State \emph{coparent} $\leftarrow$ \emph{Triangle} with index \emph{parent.id} + 1\\
		
		\LineComment{Prevent creation if either vertex ($v_e$ or $v_a$) is closed}
		\If{ \emph{parent.counters}[$v_e]=0$ \\
			\hspace*{2em} \textbf{or} \emph{parent.counters}[$v_a]=0$}
		\State \Return
		\EndIf\\
		
		\LineComment{Create new triangle}
		\State \textbf{new} \emph{child}   $\leftarrow$  \textsc{copy}(\emph{parent}) \\
		
		\LineComment{\textit{Adjust child parameter}}
		\State \emph{child.flags}[\emph{$v_n$}] $\leftarrow$ \texttt{None}
		
		\State decrement \emph{child.counters}[$v_e$] and \emph{child.counters}[$v_a$] 
		\State \emph{child.counters}[$v_n$] $\leftarrow$ \emph{child.edges}[$v_n$]\\
		
		\LineComment{Handle short branch of asymmetric filler vertex}
		\If{\textsc{vertex\_will\_be\_closed\_by\_asym}(\emph{child}, $v_e$)}
		\State decrement \emph{child.counters}[\emph{$v_e$}]
		\State \emph{child.flags}[\emph{$v_e$}] $\leftarrow$ \texttt{f} \\
		
		\If{\textsc{coparent\_is\_filler}(\emph{coparent})}
		\State \emph{child.counters}[$v_n$] $\leftarrow$ \emph{coparent.counters}[0]
		\EndIf\\
		
		\LineCommentOut{Handle long branch of asymmetric filler vertex}{\algorithmicindent}
		\ElsIf{\textsc{is\_asymmetric}(\emph{child})}
				
		\State \emph{child.flags}[\emph{$v_n$}] $\leftarrow$ \texttt{F}
		\EndIf\\
		
		\LineComment{Handle symmetric filler triangles}
		\If{
			\textsc{is\_symmetric}(\emph{child}) 
		}
		\State decrement \emph{counters}[\emph{$v_n$}]
		\EndIf\\
		
		\LineComment{Handle regular triangles}
		\If{\textsc{is\_regular}(\emph{child})}
		\State \emph{child.flags}[\emph{$v_n$}] $\leftarrow$ \texttt{F}
		\EndIf\\
		
		\State \textsc{correct\_orientation}(\emph{child}, $v_e$)
		
		\State \Return \emph{child}
		\EndFunction
	\end{algorithmic}
	\vspace{3mm}\hrule height \phaserulewidth \vspace{2mm}
	\caption{Triangle propagation mechanism, reflects a triangle on the edge spanned by $v_e$ and $v_a$. 
	}
	\label{alg:propagate}
\end{figure}

Initially, the three vertices of the parent triangle are assigned to variables $v_e, v_a$, and $v_n$.  $v_e$ and $v_a$ represent the endpoints of the reflection edge, while $v_n$ represents the non-participating vertex. Next, it is checked whether the vertices $v_e$ and $v_a$ are already closed. If one of the vertices is closed, the associated edge is considered blocked and no further triangle will be generated. If the edge is not blocked, a child triangle is created as a copy of the parent, and its attributes are adjusted. This includes updating the valence counters for the vertices $v_e$ and $v_a$. As $v_n$ is not involved in the reflection process, it will open a new vertex and thus its counter is set to the amount of triangles required at that vertex ($p$, $q$ or $r$).

Now, in a series of conditionals, the type of the child triangle is determined, using helper functions listed in Figure~\ref{alg:helper_functions}. If the asymmetric filler flag (i.e.~\texttt{f} or \texttt{F}) is set on $v_e$, indicating that this vertex will eventually be closed by an asymmetric filler triangle\footnote{Note that only for \texttt{F} the triangle is an asymmetric filler itself. For the \texttt{f} flag, the triangle is a co-parent.}.
We have chosen to close vertices with asymmetric filler triangles in a clockwise direction\footnote{The opposite choice (counter-clockwise) is equally valid.}. Therefore, we change the \texttt{F} flag on the $v_e$ vertex into an \texttt{f} flag and decrement the corresponding counter again. Moreover, if, additionally, the co-parent, given by the triangle which has been created directly after the parent in the same layer, is a filler polygon itself\footnote{Closing a vertex refers to a counter becoming zero. As the co-parent has a reduced counter for its asymmetric filler 'child', it is involved in closing this vertex. However, as in its layer the vertex is not closed yet, the co-parent is not necessarily a filler polygon. Broadly speaking, a co-parent closing two vertices is both, co-parent and filler polygon.}, $v_n$ will not open a new vertex since this vertex already belongs to a vertex opened by the co-parent. Therefore, the $v_n$ counter is decremented.

As symmetric filler triangles appear in pairs, the rotation direction (clockwise or counter clockwise) and hence the flag values are unimportant in this case. Therefore, if either $v_e$ or $v_a$ is a symmetric filler triangle, the newly created vertex at $v_n$ will be shared among the symmetric pair and thus the counter for $v_n$ is decremented. Finally, if the triangle is neither an asymmetric nor a symmetric filler, the newly opened vertex will be closed by an asymmetric filler triangle and thus the \texttt{F} flag is set on $v_n$.

In Fig.~\ref{alg:graph}, we present the pseudo code for the neighbor registration function, which operates in $\mathcal{O}(1)$ time. This complexity stems from the fact that the polygon types derived by \textsc{propagate} combined with knowledge about the parent together encode the layer structure. It is crucial to note that the polygon type itself does not encode parent-child relationships. Consequently, this function is designed to be invoked during the construction of the tiling (compare  Figure~\ref{alg:main}), rather than on a fully constructed one.

\begin{figure}[b!]
	\hrule height \phaserulewidth \vspace{3mm}
	\begin{algorithmic}
		\Function{GRAPH}{\emph{t: Triangle}, \emph{parent: Triangle}}
		
		\State \textsc{add\_neighbours}(\emph{t.id}, \emph{parent.id})\\
		
		
		\If{\textsc{is\_asymmetric}(\emph{t})}
		\State \textsc{add\_neighbours}(\emph{t.id}, \emph{parent.id}+1)
		\ElsIf{\textsc{is\_symmetric\_cc}(\emph{t})}
		\State \textsc{add\_neighbours}(\emph{t.id}, \emph{t.id}$-1$)
		\EndIf
		\EndFunction
	\end{algorithmic}
	\vspace{3mm}\hrule height \phaserulewidth \vspace{2mm}
	\caption{Code fragment establishing the core logic of neighbor registration, omitting a number of implementation details such as cyclic layer closure and boundary case handling. \textsc{add\_neighbours} is a generic method collection adjacency relations in a suitable data container.}
	\label{alg:graph}
\end{figure}
\section{Regular Polygons}
\label{sec:RegularPolygons}
Regular polygonal tilings can be obtained by the coalescing $2p$ triangles within a $(p,q,2)$ triangle group tiling. However, we also introduce a specialization of our algorithm, together with a numerical implementation (GRC, detailed in Section~\ref{sec:Performance}), capable of generating regular $(p,q)$ tilings directly. This appendix outlines the required specific algorithmic adjustments compared to the triangle case.

In the regular polygonal case,  the number of cells (polygons) sharing a vertex is identical for every vertex and the propagation method can be simplified as follows:
\begin{itemize}
	\item Compared to arbitrary triangles, regular polygons do not require storing an array of counters. Instead, a single global value, \( q / 2 \) for even \( q \) and \( (q+1) / 2 \) for odd \( q \), is sufficient.
	
	\item As long as \( q > 3 \), a single polygon cannot simultaneously close two vertices. Consequently, instead of potentially different polygon types (regular, asymmetric filler or symmetric filler) at each vertex (see helper functions in Figure~\ref{alg:propagate}), the polygon can be treated as a single, locally consistent type. As a consequence, the  special case involving the co-parent in Figure~\ref{alg:propagate} (lines 25) can be omitted. Furthermore, for $q = 3$, each polygon closes two vertices with its left and right siblings, respectively. Additionally, in case the polygon is an asymmetric filler, another vertex, shared with both the parent and the co-parent is closed.
	
	\item For $q > 3$, only the first and last child of a polygon can close an open vertex. All other child polygons are regular.
	\item Only the first and last vertex of a polygon already exist upon its creation and therefore require a counter variable. All other vertices are newly opened by this polygon. We can hence reduce the number of counters, by letting each child created on those vertices inherit a counter for that vertex -- initialized with a global default value $q/2-1$ for even $q$ and $(q-1)/2$ for odd $q$. Providing an identical counter in the parent object would hence redundant and can be omitted.
\end{itemize}
We emphasize that this specialized variant of the propagation function works for arbitrary values of $p$. The number of calls to this function needs to be adjusted from $2$ (in the triangle case) to $p-1$ for regular $p$-gonal tilings. One edge is already blocked by the parent.

\section{Formalization of the propagation rules}

This section presents analytical expressions that govern the cell propagation within our algorithm for polygonal tilings. We define $\Delta$ as the number of additional tiling layers required to close a newly opened vertex for regular $(p, q)$ tilings. 
It is given by
\begin{equation}
	\Delta = \left\{
	\begin{array}{ll}
		\frac{q - 1}{2} & \text{for } q \text{ is odd} \\
		\frac{q}{2} & \text{for } q \text{ is even} \\
	\end{array}
	\right.
\end{equation}

\subsection{Basic rules}

The polygon propagation mechanics detailed in this manuscript can be formalized as follows:

\begin{itemize}
	\item \textbf{Regular} polygons
	\begin{itemize}
		\item create $p - 1$ children for the next layer
		\item open $p - 2$ new vertices that will be closed in the ($n + \Delta)$-th layer by symmetric fillers for odd $q$ and by asymmetric fillers for even $q$
	\end{itemize}
	
	\item \textbf{Asymmetric filler} polygon
	\begin{itemize}
		\item create $p - 2$ children in the next layer
		\item open $p - 3$ new vertices that will be closed in the ($n + \Delta)$-th layer by symmetric fillers for odd $q$ and by asymmetric fillers for even $q$
		\item only asymmetric filler polygons have two parents (the actual parent and the co-parent). Since the actual parent is the one that created the polygon, the corresponding edge of the co-parent must be blocked in order to avoid duplication. In other words, this suppresses the generation of one other polygon in the current polygons layer.
	\end{itemize}
	
	\item \textbf{Symmetric filler} polygons (odd $q$ only)
	\begin{itemize}
		\item create $p - 2$ children in the next layer
		\item open $p - 3$ new vertices that will be closed in the ($n + \Delta)$-th layer by a pair of symmetric fillers
		\item each pair of symmetric filler polygons creates an asymmetric filler polygon in the $(n + \Delta)$-th layer.
	\end{itemize}
\end{itemize}

These deduction rules allow to establish a propagation formula for the number of polygons per layer.
Let $n \in \mathbb{N}$ be the label of the layer. Moreover, let $\#A_{n}$ and $\#S_{n}$ be the number of asymmetric and symmetric filler polygons in the $n$-th layer, respectively. Therefore, the total number of polygons in the $n$-th layer is given as
\begin{equation}
	\label{eq:q=odd:all_polys_per_layer}
	\#N_n = (p - 1) \cdot \#N_{n - 1} - \#S_{n-1} - \#A_{n} - \#A_{n - 1}, 
\end{equation}
\begin{equation}
	\text{with} \quad \forall{n < \Delta:\,\,}\#A_{n} = \#S_{n} = 0. 
\end{equation}

\subsection{Even $q$}

For $q$ even, the tiling does not contain symmetric filler polygons, i.e. 
\begin{equation}
	\forall{n \in \mathbb{N}:}\,\,\#S_{n} = 0,
\end{equation}
as the vertices of the fundamental polygon and its pairwise descendants will be closed by a single polygon (an asymmetric filler). The number of asymmetric filler polygons is given as
\begin{equation}
	\label{eq:k=even:first_order}
	\#A_n = (p - 2) \cdot \left[ \#N_{n - \Delta} - \#A_{n-\Delta} \right] + (p - 3) \cdot \#A_{n-\Delta}, 
\end{equation}
where $\left[ \#N_{n - \Delta} - \#A_{n-\Delta} \right]$ is the number of regular polygons in the $(n - \Delta)$-th layer, spawning $p - 2$ asymmetric filler polygons.

\subsection{Odd $q$}
For $q$ odd, the different types of fillers are created in alternating manner as each pair of symmetric filler polygons creates an asymmetric filler polygon in the ($n + \Delta)$-th layer and vice versa. Therefore, the number of asymmetric filler polygons is given as
\begin{equation}
	\label{eq:q=odd:1st}
	\#A_n = \frac{1}{2}\#S_{n-\Delta},
\end{equation} 
and the number of symmetric filler polygons as
\begin{equation}
	\label{eq:k=odd:second_order}
	\begin{aligned}
		\#S_n =& 2 \cdot \Big((p - 2) \cdot \left[ \#N_{n - \Delta} - \#S_{n-\Delta} - \#A_{n - \Delta}\right]\\ &+ (p - 3) \cdot \left[\#S_{n-\Delta} + \#A_{n - \Delta}\right]\Big).
	\end{aligned}
\end{equation}
Here $\left[ \#N_{n - \Delta} - \#S_{n-\Delta} - \#A_{n - \Delta}\right]$ denotes the number of regular polygons and $\left[\#S_{n-\Delta} + \#A_{n - \Delta}\right]$ the total number of filler polygons (both types).

\subsection{Special case $q=3$}

A special case is $q = 3$, where all polygons are positioned adjacent to their \textit{siblings} and thus close at least two vertices. A \emph{sibling} is any other cell created by the parent or a co-parent, which resides in the same layer as the considered cell. The modified propagation behavior of the polygons in the $n$-th layer reads:
\begin{itemize}
	\item \textbf{Regular} polygons:
	\begin{itemize}
		\item create $p - 3$ \textit{children} for the next layer
		\item create one asymmetric filler polygon with both of their \textit{siblings}, respectively. However each polygon is a co-parent once and thus only creates a single asymmetric filler polygon.
	\end{itemize}
	\item \textbf{Asymmetric filler polygons}:
	\begin{itemize}
		\item create $p - 2$ \textit{children} in the next layer
		\item create one asymmetric filler polygon with both of its \textit{siblings} each. However each polygon is a co-parent once and thus only creates a single asymmetric filler.
	\end{itemize}
\end{itemize}
Moreover, in the $n$-th layer, one polygon less will be created for each asymmetric filler.

The number of regular polygons is given as
\begin{equation}
	\label{eq:q=3:all_polys_per_layer}
	\#N_n = (p - 3) \cdot \#N_{n-1} - \#A_{n-1} - \#A_{n}.
\end{equation}
For asymmetric filler follows
\begin{equation}
	\label{eq:q=3:1st}
	\#A_{n} = \#N_{n-1}.
\end{equation}
Inserting Equation~\ref{eq:q=3:1st} into Equation~\ref{eq:q=3:all_polys_per_layer} yields
\begin{equation}
	\begin{aligned}
		\#N_n &= (p - 1) \cdot \#N_{n-1} - \#N_{n-2} - \#N_{n-1}\\
		&= (p - 2) \cdot \#N_{n-1} - \#N_{n-2}\\
	\end{aligned}
\end{equation}
With $k = p - 3$ and $q - 2 = 3 - 2 = 1$, it follows
\begin{equation}
	\begin{aligned}
		\label{eq:q=3:final_polyforumula}
		\#N_n &= (p - 3 - 1) \cdot \#N_{n-1} - \#N_{n-2}\\
		&= \big((q - 2)(p - 2) - 2\big) \cdot \#N_{n-1} - \#N_{n-2}.
	\end{aligned}
\end{equation}
As expected Equation~\ref{eq:q=3:final_polyforumula} matches with the formula for layer sizes used by \cite{PhysRevE.96.042116} as our layer definition matches theirs in case of $q = 3$.

\begin{figure}[t]
	\hrule height \phaserulewidth \vspace{3mm}
	\begin{algorithmic}
		\LineComment{Check if triangle is counter-clockw.~instance of a symmetric pair}
		\Function{is\_symmetric\_cc}{\emph{t}}
		\State \Return \emph{t.flags}[1] \textbf{not in} [\texttt{f}, \texttt{F}] \textbf{and} \emph{t.counters}[1] = 0
		\EndFunction\\
		
		\LineComment{Check whether triangle is clockwise instance of a symmetric pair}
		\Function{is\_symmetric\_c}{\emph{t}}
		\State \Return \emph{t.flags}[2] \textbf{not in} [\texttt{f}, \texttt{F}] \textbf{and} \emph{t.counters}[2] = 0
		\EndFunction\\
		
		\LineComment{Check whether triangle is any instance of symmetric pair}
		\Function{is\_symmetric}{\emph{t}}
		\State \Return \textsc{is\_symmetric\_cc}(\emph{t}) \textbf{or} \textsc{is\_symmetric\_c}(\emph{t})
		\EndFunction\\
		
		\LineComment{Check whether triangle type is asymmetric}
		\Function{is\_asymmetric}{\emph{t}}
		\State \Return \emph{t.flags}[2] = \texttt{F} \textbf{and} \emph{t.counters}[2] = 0
		\EndFunction\\

		\LineComment{Check whether triangle type is regular}
		\Function{is\_regular}{\emph{t}}
		\State \Return \textbf{not} (\textsc{is\_symmetric}(\emph{t}) \textbf{or} \textsc{is\_asymmetric}(\emph{t}))
		\EndFunction\\

		\LineComment{Check whether vertex $v$ will be closed by an asymmetric triangle}
		\Function{vertex\_will\_be\_closed\_by\_asym}{\emph{t}, \emph{v}}
		\State \Return \emph{t.flags}[\emph{v}] = \texttt{F}
		\EndFunction\\
		
		\LineComment{Check whether triangle $t$ closes two vertices of the tiling graph}
		\Function{coparent\_is\_filler}{\emph{t}}
		
		\State \Return \emph{t.counters}[1] = 0 \textbf{and} \emph{t.counters}[2] = 0
		\EndFunction\\
		
		\LineComment{Correct the order in which counters, edges, flags, and coordinates are stored in the array}
		\Function{correct\_orientation}{\emph{child}, $v_e$}
		\State \emph{child.counters} = \textsc{roll}(\textsc{flip}(child.counters), $v_e + 2$)
		\State \emph{child.edges} = \textsc{roll}(\textsc{flip}(child.edges), $v_e + 2$)
		\State \emph{child.flags} = \textsc{roll}(\textsc{flip}(child.flags), $v_e + 2$)
		\EndFunction
	\end{algorithmic}
	\vspace{3mm}\hrule height \phaserulewidth \vspace{2mm}
	\caption{Helpers used in methods \textsc{propagate} and \textsc{graph}.}
	\label{alg:helper_functions}
\end{figure}

\subsection{Comparison of different cases}

Despite the seemingly distinct forms of the derived equations for the various cell types, they fundamentally describe the same propagation behavior and are, in fact, mathematically equivalent when taking into account the particular geometric realizations of these types. The major differences between both cases is encoded in the propagation behavior of the filler polygons. As for $q$ is even only asymmetric filler polygons exist, the number of asymmetric filler polygons depends on the number of regular and asymmetric filler polygons only as can be seen in Equation~\ref{eq:k=even:first_order}. In contrast, for odd $q$ both types of filler polygons exist. In fact, the number of asymmetric filler polygons merely depends on the number of symmetric filler polygons as shown in Equation~\ref{eq:q=odd:1st}. However, inserting Equation~\ref{eq:k=odd:second_order}, i.e. the number of symmetric filler polygons, into Equation~\ref{eq:q=odd:1st} reads
\begin{equation}
	\begin{aligned}
		\#A_n = &(p - 2) \cdot \left[ \#N_{n - 2\Delta} - \#S_{n-2\Delta} - \#A_{n-2\Delta} \right]\\ + &(p - 3) \cdot \left(\#S_{n-2\Delta} + \#A_{n-2\Delta} \right),
	\end{aligned}
\end{equation}
which is, again, identical to the $q$ is even case shown in Equation~\ref{eq:k=even:first_order} when $\#S_{n-2\Delta} = 0$ and accounting for the alternating production of symmetric and asymmetric filler polygon such that $2\Delta \rightarrow \Delta$.

For the special case of $q=3$, the total number of polygons is described by Equation~\ref{eq:q=3:all_polys_per_layer}. Considering that for $q=3$ every cell acts as two symmetric filler polygons such that $\#N_n \rightarrow \#N_n - 2\#S_n$ and considering the different children creation rates of $p - 3$ and $p - 1$ for $q=3$ and $q$ is odd, respectively, Equation~\ref{eq:q=3:all_polys_per_layer} and Equation~\ref{eq:q=odd:all_polys_per_layer} are in fact identical again. Similarly, Equation~\ref{eq:q=odd:1st} and Equation~\ref{eq:q=3:1st} are identical considering $\#N_n \equiv 2\#S_n$.

\section{Memory efficiency}
\label{sec:AppendixMemory}

\begin{figure}[t!]
	\includegraphics[width=0.9\linewidth]{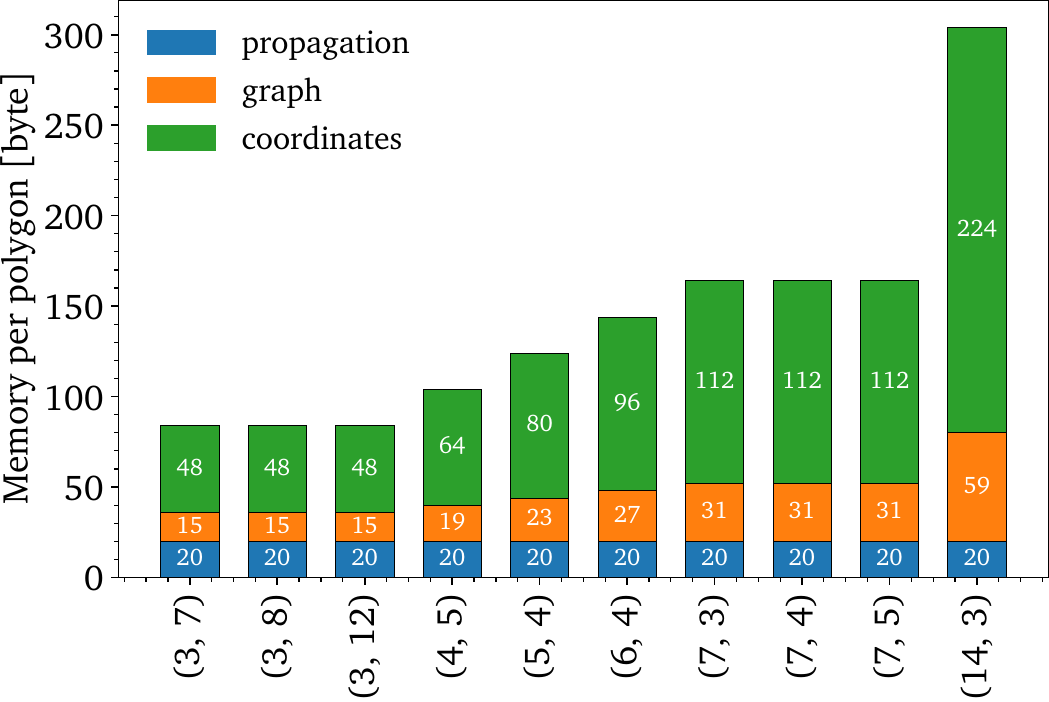}
	\caption{Memory requirement for the GRC kernel separated into propagation, graph and coordinate shares.}
	\label{fig:app_memory_bench_full}
\end{figure}

In the following we examine the memory requirements of the GRC kernel implementation as measured using the \textit{tracemallloc} Python library. Results are shown in Figure~\ref{fig:app_memory_bench_full}. When using coordinates, the memory footprint is typically dominated by them, as for each vertex a 128-bit (16-byte) complex numbers (given the Poincare disk representation) is stored, leading to a memory complexity of $\mathcal{O}(p)$ per polygon. Accordingly, in Figure~\ref{fig:app_memory_bench_full}, the green bars (``coordinates'') account for $16p$ byte.

The neighbor relations are stored as 32-bit integers (4 bytes), again yielding $\mathcal{O}(p)$ memory complexity. Technically, an additional integer counting the number of neighbors is required, rendering the footprint for orange bars (``graph'') as $4(p + 1)$. In the figure, we find a systematic underestimation by 1 byte, which we assume to be an artifact.

Finally, the memory footprint attributed to the propagation of cells is constant. This is expected, as for each polygon only the type as well as the states of the left-most and right-most vertices are stored, which makes it independent of both $p$ and $q$. 

\bibliography{literature.bib}
\end{document}